# CONCEPT OF DELIVERY SYSTEM IN THE SMART CITY ENVIRONMENT

ZuzanaŠpitálová, Oliver Leontiev and PatrikHarmaňoš

Institute of Computer Engineering and Applied Informatics Faculty of Informatics and Information Technology, Slovak University of Technology, Bratislava, Slovakia

*ABSTRACT*

*Regarding to the smart city infrastructures, there is a demand for big data processing and its further usage. This data can be gained by various means. There are many IoT devices in the city, which can communicate and share the information about the environment which they are situated in. Moreover every personal mobile device can also participate in this process and help to gain data via various applications. Every app provides the useful data, enabling the location and data sharing. This data can be further processed and used for improving the city infrastructure, transport or other services. We designed the system for shared delivery process, which can help to achieve the described situation. It consists of frontend and backend part. The frontend part, multiplatform mobile app, represents the graphical interface and the backend part represents the database for the gained data.*

## 1. INTRODUCTION

More than a decade, the technical progress of the city infrastructure is increasing in many ways. Using the complex technologies, there is a contribution to smart city environment creation. Smart city can be defined as the place, which uses the modern technical solutions for conventional services and networks for improving the daily life of the people which lives there [1]. The information and communication technologies (ICTs), like artificial intelligence or autonomous vehicles, are involved into daily aspects of life for supporting the urban development. The important item of smart city environment is Internet of Things (IoT) [2]. It represents the devices, which are connected and able to communicate between each other. Via these devices, the information can be gained and processed for other purposes.

Based on the mentioned definition and used ICTs, there is a city ranking, which evaluates the achieved level of smart city. According to [3], these studies are done regularly to find the smartest city of the year. There are many aspects, which are required for achieving the certain level of smart city. There is rated the integration of urban technologies, sensors and any personal device. Via them, the useful data can be gained and further used for improving the existed services or finding the information. The data can be also gained by various applications used in the city, like tourist apps, traffic apps, etc.

There is another study related to the citizen cloud, implemented in the city of Shanghai [4]. There can be seen the implementation of the city platform, which provides the public services, like culture or healthcare, transportation and etc. So, all the information, related to the services in the city, can be found there.

In the last years, there is an increasing number of mobile apps, which use the device location during providing the required service. This location can be seen on the digital map. All these devices, via used applications, can provide the data related to the environment, which they are situated in. Using this gained data, there is much useful information. Between these apps, there are services like taxi, food delivery, shared travelling, persons and goods transport. As there is an intention to provide the services via apps, we designed the system for shared delivery, consisted of frontend and backend part. We studied also the systems for delivery process, which are available on the market. The most of them belongs to the portfolio of local and global companies, like post and courier services. There was the possibility to study only the frontend part of them. These apps are used provided that, there is the courier, which is employee of the company, and the customer, which receives or sends a packet. The main idea of our system is the shared delivery. So, everyone can participate in the process choosing the courier or user mode. We suppose the shared delivery in the cases, when the delivery person and the delivered parcel have the same destination address. This helps to save sources and decrease emissions. The other characteristic of our system is the possibility to gain the data from the environment and its further usage. By the conventional systems, belonged to the big delivery companies, there is not the information about this functionality.

Regarding to the system's design forshared delivery in smart city environment, the used technology has a significant importance. The main aim is the multiplatform usage. Thanks to this possibility, many people can participate in that process independently from the mobile platform.
To achieve the multiplatform usage, the React Native [5] is a good choice for frontend part of the system. It is open-source framework based on JavaScript for mobile apps development using the components. As it is open-source, it is not needed to develop everything by yourself. As the frontend part represents the graphical interface, the other characteristic is the user-friendly functionality and usage. There is also the request for quick and secure communication with the server and the secure local data storage.

Regarding to the backend part of the system, the multiplatform characteristic of the database system has also the importance. We chose the PostgreSQL [6] database because of the Relational Database Management System (RDBMS) requirement to provide the relational database management functionality. The data is stored in the tables. RDBMS ensures the integrity and rules between databases. The first of the main requirements for the backend part is the speed of the system. So, the server has to respond as quickly as possible. The second requirement is the encryption, related to the communication with the frontend part, and the data which is stored in the database.

## 2. DESIGN

As it was mentioned before, the aim of our work was to design the system for shared delivery process, which could be used by people in the certain urban area, consisted of frontend and backend part. The main idea is the shared delivery in case of the same direction of delivered parcel and travelling person. So, everyone in the community can participate in the process. The means of transport (car, van, motorcycle, e-bike, etc.), used for delivery, depends on the person, which wants to use the proposed system. We wanted to create the open-source project, which could be used by various people, delivering the items during the travelling to school, work, etc., if the destination address of the item would be the same as they intend to reach. This helps to save time, money and also the other sources. As the smart city environment is also ranked according to the implementation by various technologies and devices used inside it, the other aim of our project is the usage in the urban area, where will be the possibility to gain various data regarded to traffic, large concentration of cars, number of daily orders, etc. The gained data can be processed and used for other purposes. Using that, there can be the infrastructure improvements, green and busy zones pick outs, and many others assets regarded to the daily life in the city.

## 2.1. Frontend

Our proposal was designed in the open-source multiplatform framework, React Native. Regarding to this, it can be used by multiple mobile platform, like iOS and Android. The app works in respecting the good manners.

The functional and non-functional requirements were also considered. From non-functional point of view, there are the following considered requirements for the app:
- Multiple mobile platform availability
- Check of the input data from the users
- Request check during the communication with server (backend)
- User interface should be able to adopt for various type of devices
- Application will send the delivery person location every few seconds
- The delivery person role will be available only under registration
- The data from the server will be provided in JSON format.
-

From functional point of view, there are two roles, the user and the delivery person. It is not necessary to use the both roles. The requirements were defined for both separately and for both in common. The common functional requirements for both roles are the following:
- Registration: User account creation.
- Login: Login into existing account via email and password.

After registration or login, there are other available possibilities. For the user role regarded to functional requirements, there are the following items:
- Delivered package tracking: The possibility to track the delivered package.
- Availability of user profile: User information, like name, email, address, etc.
- Edit the user profile: The possibility to change the user data.
- Add the payment data: The data regarded to credit card. This item is optional.
- New item creation: Delivered parcel creation and specification.
- My items: Active and inactive items. It means the delivered packages and the ones in delivery.
- Statistics: Information about the activity.
- Delivery person location: Availability of updated package location during the delivery process.

For the delivery person role, respecting the functional requirements, there are the items:
- Registration as delivery person: After creation the account, the user role is automatically chosen. For the delivery person role, it is necessary to choose this role in the app's menu.
- Active items: Availability of the item in the delivery.
- Acceptance of parcel for delivery: The item for accepting the package which can be delivered by the delivery person.
- Parcel delivery: Mark the end of the delivery by courier.
- History of delivered items.

For the usage of app functionalities, it is necessary to communicate with the server (backend) part of our project. The communication is realized via the REST API interface using HTTPS protocol. REST API enables the data transfer between the app and the server. HTTP request has to be authorized. The access token is necessary for the authorization. It is gained during the login. After login, the server sends the access token and the renew token, which are stored in the local storage of the mobile device. So, the access token is used in every request. If the access token is not valid, the renew token is used. If this one is also not valid, the server sends an error and the user has to login again. If the renew token is valid, the server sends the new access and renew one, which are again stored in the local storage. Although the application communicates with the server and most of the gained data are stored there, there are still the situations when some data needs to be stored locally. This can happen when there is no internet connection or connection failure. If the data is stored locally, the request sent to the server is not necessary. For that purpose, the global data storage was created in the app. It uses React Context API, which represents the interface enabling the data transfer between components. The data can be stored in two ways, using Async Storage and Secure Storage. Async Storage is used in case of non-sensitive data. It is the React Native system maintained by the developer's community, providing the local unencrypted asynchronous key-value storage. Secure Storage enables to store the sensitive data. The platform iOS supports the iOS-Keychain Services storage for certificates or tokens. The platform Android stores the data in key-value form using Secure Shared Preferences with Android Keystore System encryption. For both systems implementation, it is necessary to use the extern library. In our work, it is Expo Secure Store.

So, the application works after establishing the communication via registration or login. The user can choose which role wants to use. In the case of standard user role, the user can be the sender and also the receiver of the package. The receiver does not need to be registered. By the sender sub-role, the registration is necessary. After registration and login, there is the item for new package delivery creation. The necessary information has to be filled to create successful order. The required data are package dimensions, address of the sender and the receiver. For filling the address, the Google Places Autocomplete component is used. It uses the Google Places API. After that, the new parcel can be seen on the map available in the application. There is also the possibility to see the available delivery persons in the surrounding area on the same map. After this process, the package is in the system waiting for the person which can deliver that. After accepting and picking up the package by delivery person, the parcel tracking is available in the app, until it is delivered to the destination address. In the case of delivery person, it is necessary to register into this role. After filling all the necessary data (name, means of transport, etc.), the person can participate in the process and the available parcels are visible for him. After accepting some parcel for delivering, the location of the delivery person is tracked and seen on the app's map. This functionality is enabled via the websockets. This websocket connection is created between the app and the server by choosing the package for delivery. The location is sent periodically every four seconds and has to be enabled on the delivery person's device. The information about the current latitude and longitude is sent to websocket server until the package is delivered.

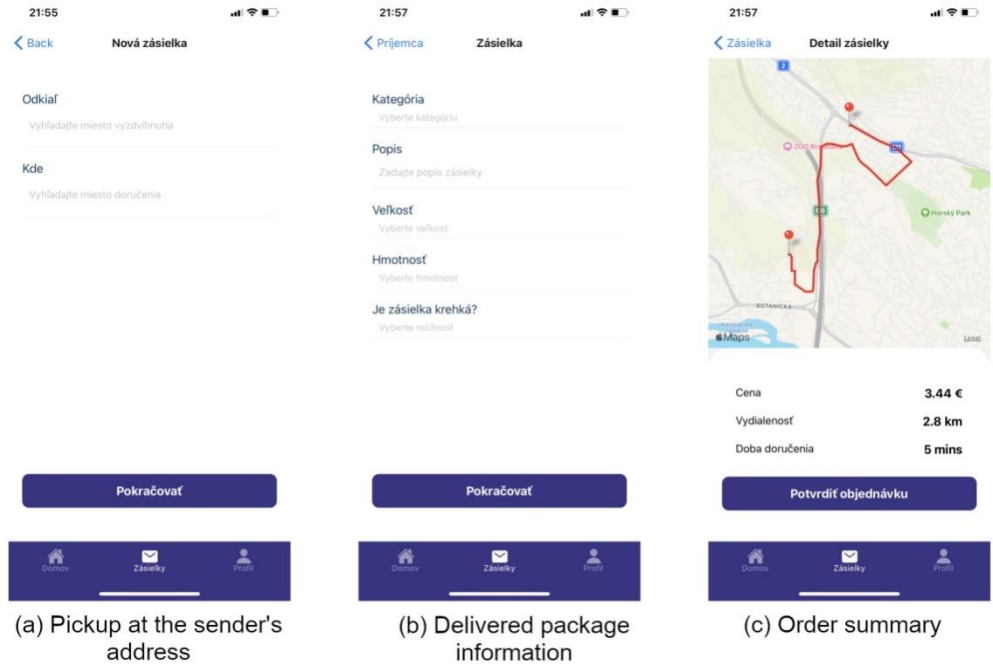

Figure 1. Delivery order creation screen

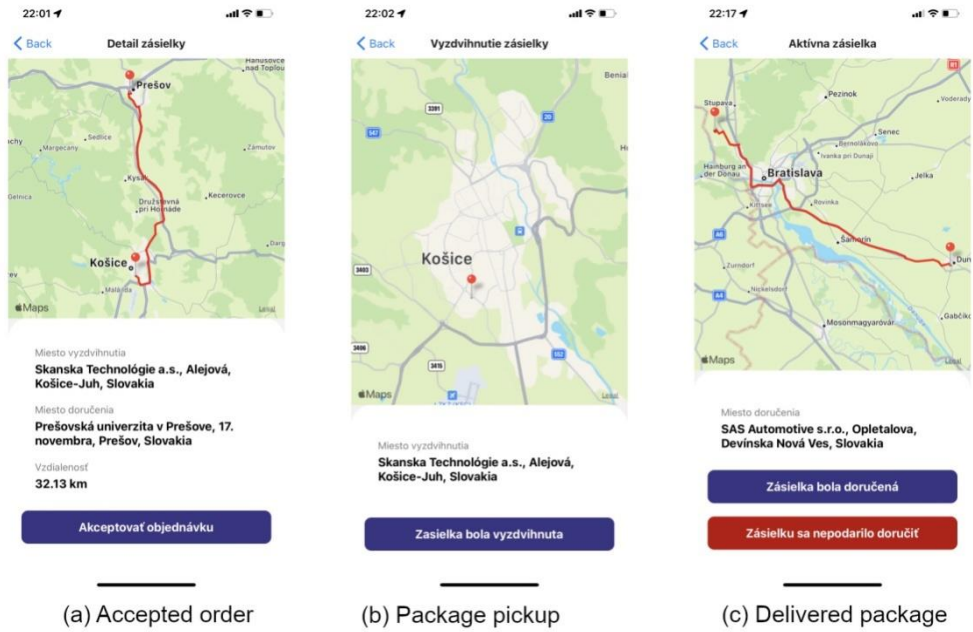

Figure 2. Package delivery screen

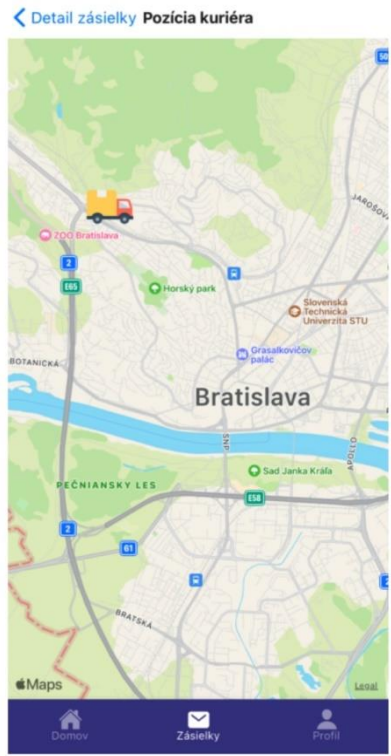

Figure 3. Screen of courier location in real time

### 2.1.1. Testing

Regarding to the testing, it was realized in two phases. The first phase was aimed to functional and non-functional requirements, which were defined during the development. The testing process was iterative. It means, every added functionality was tested separately first. If the first test results were acceptable, the new functionality was also tested with the others, which had been already implemented before. If the first test was not acceptable, the review process was started and the designed functionality was refactored.

The second phase was aimed to users' testing. This means, the app was tested regarding to user interface and app functionality by real users. There were five metrics used for the experiment. The first metrics is the success of tasks fulfillment, which was achieved by the user. The second is the number of identified failures in the system observed by the user. The third is the time needed for tests execution. The forth represents the unnecessary steps realized during the testing in comparison with the ideal way. The last one is the questionnaire about the system usage based on SUS (System Usability Scale) method [7, 8]. There are 10 questions for ranking the application on a scale from 1 to 5. On a scale, 1 means strongly disagree, 2 is disagree, 3 is neutral, 4 means agree and 5 is strongly agree. According to [7, 8], there is a calculation and result can be ranked from A to F. A means the best result and F is the worst one. The system was tested by 5 users.

Regarding to the first metrics, four of the users fulfilled all tasks. One of them was not able to find and create the delivery person account.

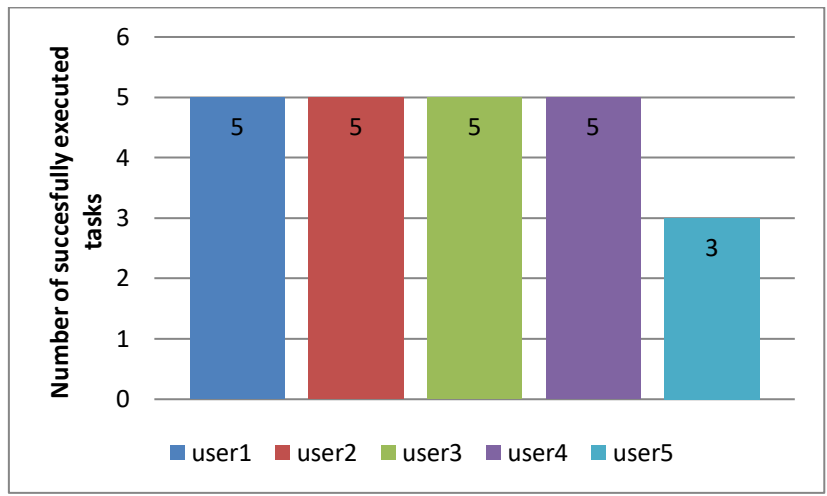

Figure 4. Success of tasks fulfillment

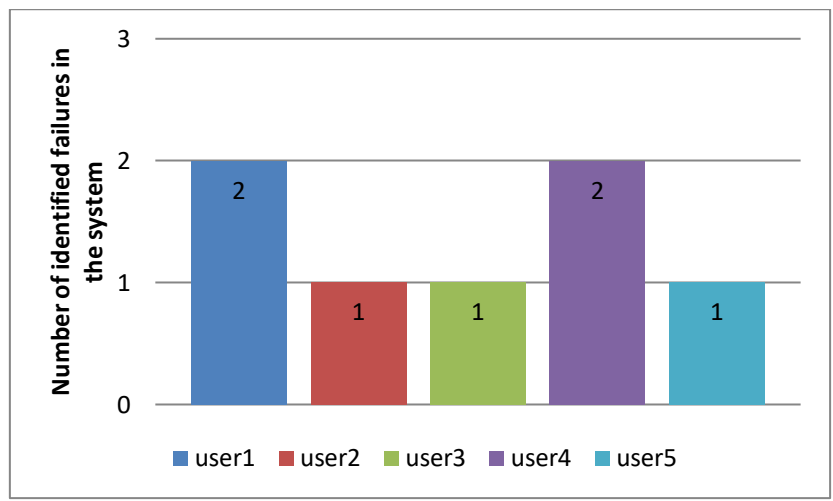

Figure 5. Identified failures in the system

By second metrics, the first user identified two issues. One issue was related to the case, when wrong email address was filled from the device memory. The problem was that the field reacted just to the manual keyboard text insert. The other issue was the date field by delivery person registration, which did not have specified format type. The second user identified the issue with Package delivery screen, where the package information overlapped the map environment. The third user identified the issue with Pick up button on the Package pick up screen. The button did not react. The problem was by Android platform, caused by missing specified field in the header of sent request. The forth user found two issues. The first issue was the password change failure and the second one was the email-address overhanging the field. The last user identified one issue related to the delivery person account creation. The user was not able to create the account. He found the process complicated. The other metrics were comparing the time and steps needed for tasks' fulfillment. For some of the users, it was difficult to find the required functionalities on the first time. They had to click between screens more time. The last metrics was evaluated by mentioned SUS method. Regarding to that, three users evaluated the proposed system as excellent (A) and two of them evaluated that as good (B).

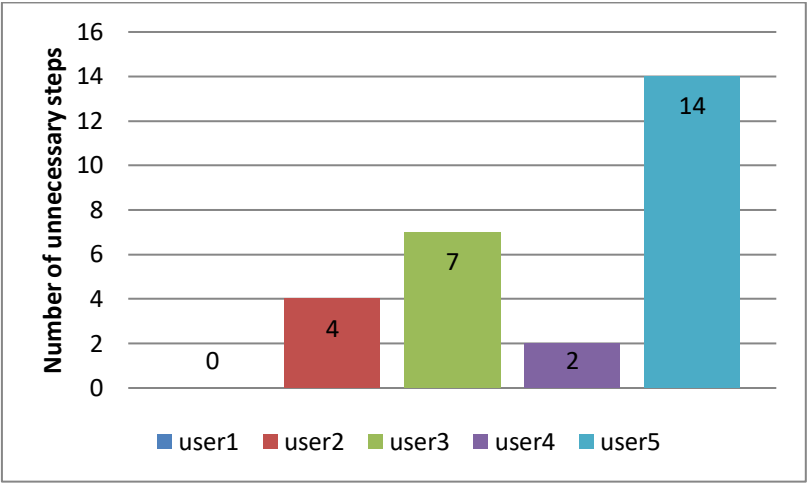

Figure 6. Unnecessary steps needed for tasks fulfillment

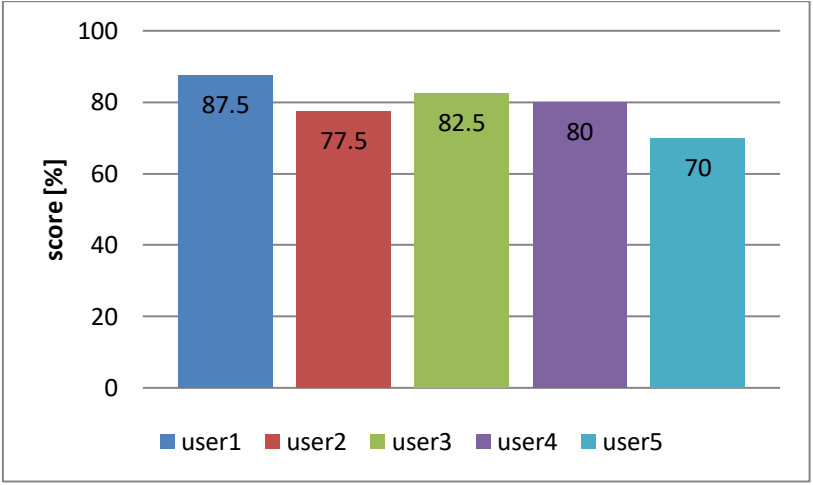

Figure 7. System Usability Scale (SUS)

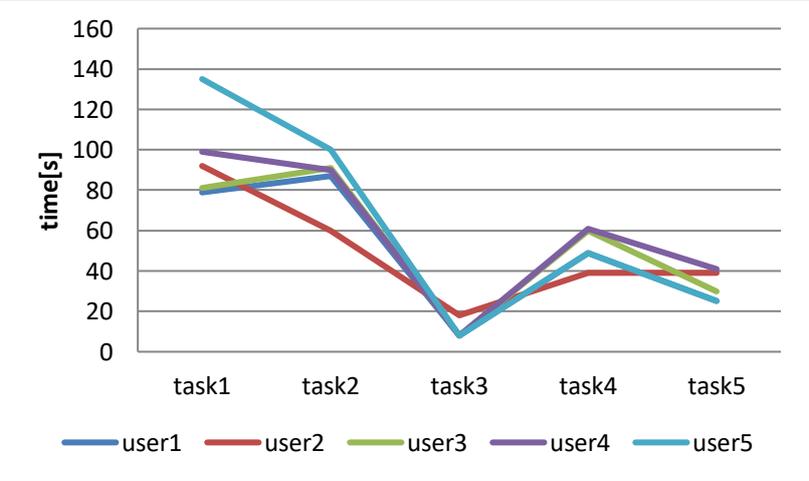

Figure 8. Time needed for tasks fulfillment

## 2.2. Backend

The backend part of our system is designed in Python with Django modules. Django is expanded with DRF (Django REST Framework), which helps to support RESTful API. We use the Nginx server and PostgreSQL database with pgcrypto and postgis modules to protect and store the encrypted and geographical data. For working with the routes between the sender and the receiver, the Google Maps API is used. The service SendGrid is used for automatic emails. For authentication via tokens, we use the JWT (JSON Web Token) technology. The protocol WebSocket is used for the communication in the real time.

For the backend part, the functional and non-functional requirements were defined. The non-functional requirements represent the characteristics of the system. We defined the following:
- Speed: The server has to respond in very small time after receiving the request.
- Encrypted communication: The communication between server and client (frontend) has to be encrypted.
- Encrypted data: Sensitive data in the database, like names, emails, etc. has to be encrypted.
- Documentation: The server interface has to be well documented for the frontend implementation purposes.
- JSON Format: Communication between client and server has to be in JSON format.

From the functional requirements point of view, there are the following items:
- Registration: The registration data has to be sent to the server to create the user's identity.
- Login: After sending the user's login data to the server, it returns the token which is used for other requests from the same user.
- Password update: Password update in the case of forgotten one.
- Package delivery creation: Item creation for delivery.
- Package delivery display: The possibility to display the packages which are delivered for the receiver. The authentication is not necessary.
- Delivered package tracking: The possibility to track the package. The authentication is not needed.
- History of packages: The list of items, which were sent.
- Delivery person registration: Registration for the users, which want to participate like the delivery persons.
- Achievement for package delivery request: The delivery person can achieve the request for package delivery in his surrounding area.
- Acceptance of package delivery request: The delivery person can accept the request.
- Change of delivery package state: The delivery person is able to change the package state. The package is delivered after marking that by this person.
- Delivery person location: The delivery person can share his location in real time.
- Access to the delivery persons' location: It enables to access the delivery persons' locations in the real time without authentication.
- Access to the delivery persons' routes: It enables to access the routes of the delivery persons, which are actually delivering without authentication and filtering.
- Access to the statistics of sent packages: Authenticated user can access the data about the sent packages.
- Email sending: Application can inform the sender and the receiver about the package.
- Administrator: Application supports the administrator interface for editing the data in the database.

The backend part is designed for two roles, the client and the delivery person (courier). The client role supports two possibilities of usage, the sender and the receiver. The sender has to be registered in the application. The data related to the sender are stored in the server using the AES [9] encrypted algorithm for private data encryption. For password encryption, the Argon2 algorithm [10] is used. After delivery request creation, the sender receives the unique code for tracking the delivered package. This code is also automatically sent to the receiver, which does not need to be registered in the app. If the receiver registers to the app, he can automatically see the delivered packages. The history of the sent parcels is available for the user. In case of delivery person role, the registration is needed. When the delivery person is active, the available packages are displayed for him according to the distance from his actual location. If the package cannot be delivered in some case, this should be communicated between the delivery person and the sender of the package.

The server part consists of data specifications, which are needed to be stored, and relations between them and APIs, which enable to work with the data. The logical model for our usage consists of various relations, like:
- Account: It represents the user's account. It contains the data about the user's role, admin and active user, and about the login data, email and password.
- Person: It contains the registration data of the users. It is also used by the delivery creation for identifying the user, which is not needed to be registered.
- Courier: It contains the information about the delivery person.
- Delivery: It is the main relation of the app. It contains the information about the delivery package state, route distance, expected delivery time, etc.
- Item: It represents the information about the delivered package, like size, weight, fragile, etc.
- Place: It represents the source and destination address of the delivered package.
- Route: It contains the data about the geographical locations of the delivery person.

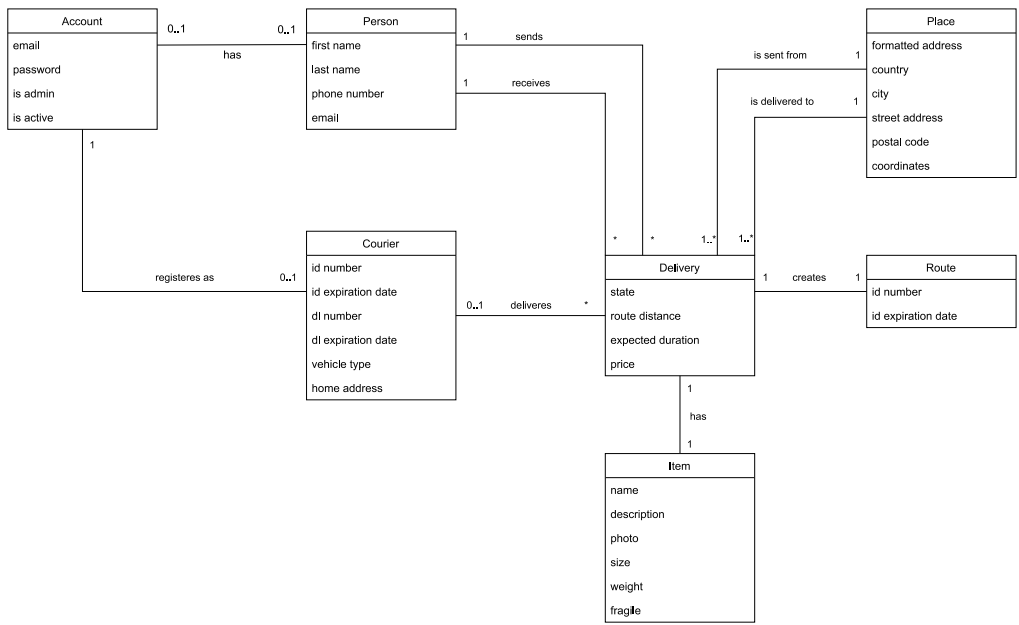

Figure 9. Logical model diagram of designed system

After consideration, we specified more precisely some of the items in the relations Courier, Delivery and Item. By the Courier relation, we specified the vehicle type with three possibilities, small, medium, large. By weight of Item relation, there was a change with possibilities, light, medium, heavy. By delivery state of Delivery relation, there are five possibilities, like ready, assigned, delivering, delivered and undeliverable.

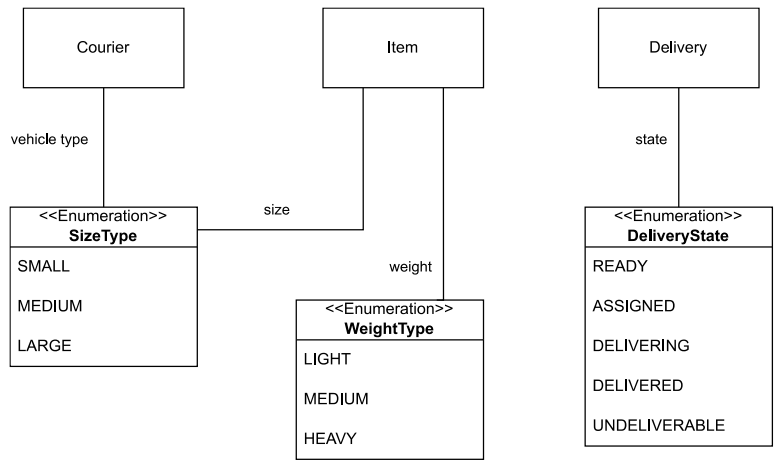

Figure 10. Enumeration diagram of designed system

The next focus is the APIs of our server application, which are in fact the URL addresses with implemented HTTP method. Respecting the principles of REST architecture, the URL addresses represent the items. The following items are the example of our designed APIs used for communication with frontend part of the system.

Table 1.Proposed APIs

| HTTP method | Final part of URL | Description |
| --- | --- | --- |
| POST | api/accounts/ | New user creation (registration) |
| POST | api/accounts/verification_email/ | Verification email re-sending |
| GET | api/accounts/token/ | Authentication user code (token) getting (login) |
| GET | api/accounts/me/ | Own account data getting |
| PATCH | api/accounts/me/ | Own account data changing |
| POST | api/accounts/reset_password/ | Password resetting |
| POST | api/deliveries/ | Delivery creation |
| GET | api/deliveries/<id>/ | Delivery data based on ID getting |
| GET | api/deliveries/ | Delivery based on history getting |
| POST | api/couriers/ | New courier creation (registration) |
| POST | api/deliveries/<id>/state/ | Delivery state change |
| GET | api/couriers/closest_delivery/ | Closest delivery getting |
| GET | api/deliveries/statistics/ | Delivery statistics getting |
| GET | api/routes/ | Routes statistics getting |

**2.2.1. Testing**

The system was tested periodically during the development, using automated testing after each functionality implementation. Besides that, the tests related to functional and non-functional requirements were realized.

Regarding to the non-functional requirements, the following items were assessed:
- Speed: The average response time of the server is 0.4 seconds.
- Encrypted communication: For encryption, it is used the HTTPS protocol and SSL certificate.
- Encrypted data: The private data, stored in the database, are encrypted using AES algorithm. For the passwords, it is used Argon2 algorithm.
- Documentation: The server API is fully documented using specification Open API 3.0.
- JSON Format: The only API, which uses the different format from the JSON type, is the package delivery creation. This uses the format Form Data due to the possibility ofadding picture into the request.

To summarize the information above, the system fulfills the non-functional requirements.
From the functional requirements point of view, there are the following results:
- Registration: The standard registration process was realized by sending the request with registration data. The verification email was sent. The account creation was also checked. The server response took 0.7 seconds on average.
- Login: Using the access token gained in the previous test, there was a check of user authetification by sending the HTTP request with GET method. The server response with private user data took 0.5 seconds on average.
- Password update: The test was realized by using the user email address, created in the first test. Then we clicked on the link for the password change. The new password was filled and it was tested for login. The login was done also with the old password. The test was successful and the server response took 0.2 seconds on average.
- Package delivery creation: The new client was created with the receiver role. The delivered package was created for the new client. We checked, if the server responded with the expected data (data related to the package) and if the information email was already sent. The test took 0.5 seconds on average.
- Delivered package tracking: The request with package ID, created in the previous test, was sent. The authetification was not done. The first step was repeated but with the authetification. After that, we observed that the server sent data about delivered package for both rounds, and only in the second round, the receiver was identified. This was the expected system behavior. The server responded in 0.1 seconds on average.
- History of packages: The other two packages with different source and destination address were created. They were sent to the test receiver. The request GET was sent.We identified us as the user, which sent the packages. The step two was repeated, but with the receiver authetification. After that, we checked the items' lists from the both rounds. The server responded in 0.2 seconds on average.
- Delivery person registration: The other account for test delivery person was created. By sending the request, we identified us as the test delivery person. We checked the expected response. The process took 0.2 seconds on average.
- Achievement for package delivery request: We sent the request, where we identified us as the delivery person, created in the previous test. We checked if the server responded with the list of items, created in the previous tests. The items should be ordered according to the distance from the delivery person location. The step one was repeated with the different location. The list of items should be in different order. The test took 0.6 seconds on average.
- Acceptance of package delivery request: We identified us as the delivery person and accepted the package for delivering, created in the previous test, sending the request to the server. We checked if the server responded with the data related to the delivered package. The state of the package and the delivery person were also changed. The process took 0.5 seconds on average.
- Change of delivery package state: We identified us as the delivery person from the previous test and checked, if the accepted packages had the changed state. The response took 0.5 seconds on average.
- Delivery person location: Using the delivered package ID, we created the connection and sent the coordinates of the delivery person. We checked if the server responded with the added delivery person ID. The test was successful with the immediate response.
- Access to the package location: Using the delivered package ID, we established the connection without authetification. The other messages were sent using the connection established in the previous test. We checked if the new messages could be seen also in the new communication. We were not able to send messages via the other connection due to the missed authetification process. The server responded correctly.
- Access to the delivery persons' location: The previous test was repeated when the connection was established as global one. The delivery persons' locations were visible.
- Access to the delivery persons' routes: The request related to the delivery persons' routes was sent and checked. The routes of created package deliveries were seen. The filter functionality was checked, too. The server response took 0.7 seconds on average.
- Access to the statistics of sent packages: We sent the request and identified us as the test sender. We checked the expected response with three created items in the last five months. The response took 0.1 seconds on average.
- Email sending: As the automatic email was already checked by package delivery creation, we changed the delivered package state to delivered.After that, we checked if the email was sent to the sender. The test was successfully done.

International Journal on Cybernetics & Informatics (IJCI) Vol. 12, No.1, February 2023

- Administrator: The administrator part was realized as fully functional.

Regarding to the functional requirements, the tests were successfully done with optimal average times. Of course, the measured times can be different on the other running environment depending on performance.

Table 2. Average request times for functional requirements

| Functional requirement test | Average request time [s] |
|---|---|
| Registration | 0.7 |
| Login | 0.5 |
| Password update | 0.2 |
| Package delivery creation | 0.5 |
| Delivered package tracking | 0.1 |
| History of packages | 0.2 |
| Delivery person registration | 0.2 |
| Achievement for package delivery request | 0.6 |
| Acceptance of package delivery request | 0.5 |
| Change of delivery package state | 0.5 |
| Delivery person location | 0 |
| Access to the package location | - |
| Access to the delivery persons' location | - |
| Access to the delivery persons' routes | 0.7 |
| Access to the statistics of sent packages | 0.1 |
| Email sending | - |
| Administrator | - |

## 3. FUTURE WORK

As we designed and tested our system as the prototype one, we plan to evaluate that by more complex testing. The tests were realized in the standard city environment. According to the data which are processed by backend, we plan to analyze the following city aspects, like:
- Traffic jams in the certain areas
- The most used means of transport for delivery (According to this, there can be prediction for infrastructure improvements, like chargers, parking places for cars, e-bikes, scooters, etc.)
- The owner of the means of transport (own, rented, shared)
- GPS and network coverage in the urban area
- Rush hours identification
- The most delivered package types (size, fragile, etc.)
- The most efficient user (private person, company)
- Etc.

We plan to make the API implementation for support various sensors (humidity, temperature, air quality), which do not require much power and do not send much data. These sensors can be mounted on users' means of transport and collect other data for processing.

The other step can be the standard improvement of our system, like security, effectiveness, database improvements for searching, whole system performance improvements, etc.

We can also extend the system to multiple servers which could be split into decentralized clusters. Implementation of this feature can improve the network stability, as the data will be stored at more places.

At the end we can try to make it open-source, as anyone can contribute and make this system better.

## 4. CONCLUSION

We designed the system, consisted of frontend and backend, for shared delivery process. Each part of the system was tested according the functional and non-functional requirements. The frontend part was also evaluated by five users, which tested the app from the user's point of view. During that, some technical failures were found and fixed. One of our testers documented also the issue related to the menu navigation, which he found complicated. We did not change the menu navigation. It has to be evaluated by more users. By backend part, there were the tests after each functionality implementation. If there was an issue, it was fixed. If unit test passed by the implemented functionality, the automated testing of whole system was executed. The time and the functional requirements were observed and documented. After the separate testing of both parts, there was also the common evaluation. As there were not so many participants in testing, we plan to do more complex evaluation process.

The main advantage of our project is the database system, which process the gained data for further purposes. According to the data analysis, there can be the improvements related to the city infrastructure and daily life.

### ACKNOWLEDGEMENT

This publication has been written thanks to the support of the Operational Programme Integrated Infrastructure for the project: Research in the SANET network and possibilities of its further use and development (ITMS code: 313011W988), co-funded by the European Regional Development Fund (ERDF).